\documentclass{article}
\usepackage{graphicx}  
\usepackage{amsmath}   
\usepackage[compress]{cite}
\usepackage{amssymb}   
\usepackage{bm} 
\usepackage{dcolumn}
\usepackage{color}
\usepackage{mathrsfs}
\usepackage{amsfonts}
\usepackage{varioref}
\usepackage{float}
\usepackage[caption=false]{subfig}
\RequirePackage[colorlinks,citecolor=blue,urlcolor=magenta,linkcolor=blue]{hyperref}
\input epsf
\usepackage{tablefootnote}
\def\gr{general relativity}

\labelformat{section}{Section #1} 
\labelformat{subsection}{Section #1} 
\labelformat{subsubsection}{Section #1}
\labelformat{subsubsubsection}{Section #1}
\labelformat{equation}{Eq.~(#1)} 
\labelformat{figure}{Fig.~#1} 
\labelformat{subfigure}{Fig.~\thefigure#1} 
\labelformat{table}{Table~#1} 
\labelformat{appendix}{Appendix #1}
\title{\bf Testing black holes in non-linear electrodynamics from the observed quasi-periodic oscillations}

\author{Indrani Banerjee\footnote{banerjeein@nitrkl.ac.in}~$^{1}$\\
{\small{$^{1}$Department of Physics and Astronomy, National Institute of Technology, Rourkela-769008, India}}}
\date{ }  
\begin{document}
  
\maketitle
\begin{abstract}
Quasi-periodic oscillations (QPOs), in particular, the ones with high frequencies, often observed in the power spectrum of black holes, are useful in understanding the nature of strong gravity since they are associated with the motion of matter in the vicinity of the black hole horizon. Interestingly, these high frequency QPOs (HFQPOs) are observed in commensurable pairs, the most common ratio being 3:2. Several theoretical models are proposed in the literature which explain the HFQPOs in terms of the orbital and epicyclic frequencies of matter rotating around the central object. Since these frequencies are sensitive to the background spacetime, the observed HFQPOs can potentially extract useful information regarding the nature of the same. In this work, we investigate the role of regular black holes with a Minkowski core, which arise in gravity coupled to non-linear electrodynamics, in explaining the HFQPOs. 
Regular black holes are particularly interesting as they provide a possible resolution to the singularity problem in \gr. We compare the model dependent QPO frequencies with the available observations of the quasi-periodic oscillations from black hole sources and perform a $\chi^2$ analysis. Our study reveals that most QPO models favor small but non-trivial values of the non-linear electrodynamics charge parameter. In particular, black holes with large values of non-linear electrodynamics charge parameter are generically disfavored by present observations related to QPOs.

\end{abstract}
\section{Introduction}\label{QPO_Intro}

The theorems of Hawking and Penrose \cite{Hawking:1973uf} state that black hole (henceforth as BH) singularities are an unavoidable feature of general relativity. However, it is believed that spacetime singularities should not exist in Nature and a suitable theory of quantum gravity can address this issue. In this regard various quantum gravity models have been put forward \cite{Horava:1995qa,Horava:1996ma,Polchinski:1998rq,Polchinski:1998rr,Ashtekar:2006rx,Ashtekar:2006uz,Ashtekar:2006wn,Kothawala:2013maa,Kothawala:2015qxa}. In the absence of a well established quantum gravity model one can address the black hole singularity problem classically by studying black holes with regular cores. Such black holes have horizons such that their metric and curvature invariants are non-singular at all points in spacetime. Motivated by quantum arguments \cite{Ansoldi:2008jw,1966JETP...22..378G,Fan:2016hvf} it was proposed that the central region of the black hole at $r\simeq 0$ should be de-Sitter like which was studied extensively \cite{Frolov:1988vj,Mukhanov:1991zn,Brandenberger:1993ef}. These works propound that quantum fluctuations prevent the unlimited increase of spacetime curvature during the stellar collapse process giving rise to black holes with regular cores.

The first regular black hole solution with a de-Sitter core was proposed by Bardeen \cite{PhysRev.174.1559} which turns out to be a solution of Einstein's equations coupled to non-linear electrodynamics. Later Ayon-Beato and Garcia \cite{Ayon-Beato:2000mjt} explained that the physical source that gives rise to the Bardeen solution corresponds to the gravitational field of a nonlinear magnetic monopole of a self-gravitating magnetic field.
Several other regular black hole solutions of gravity coupled to non-linear electrodynamics have been obtained later \cite{Ayon-Beato:2000mjt,Ayon-Beato:1998hmi,Ayon-Beato:2004ywd}. Since then there has been a lot of research in the direction of investigating regular black holes \cite{Bronnikov:2000vy,Borde:1996df,Barrabes:1995nk,Ayon-Beato:1999qin,Bonanno:2000ep,Nicolini:2005vd,Myung:2007qt,PhysRevLett.96.031103}.

In this work, we study a class of regular black hole in non-linear electrodynamics with an asymptotically Minkowski core. Such a metric arises as the solution of Einstein's equations with an anisotropic fluid resembling the Maxwell stress tensor far from the source \cite{Culetu:2014lca}. The electric field from such a source asymptotically resembles the Coulomb field, is bound at all distances, and the corresponding geometry is like that of Reissner-Nordstrom (RN) metric far from the source. The motivation for exploring such black hole solution stems from the fact that the mass function has an exponential convergence factor which makes the quantum gravity model finite to all orders upto the Planck scale \cite{Brown:1980uk}. Further, study of a finite quantum gravity is important as it can address the cosmological constant problem \cite{Moffat:2001jf} and can eliminate the divergences arising in flat space quantum field theories.

Since black holes observed in nature are in general rotating the stationary and axi-symmetric counterpart of the aforesaid regular solution with Minkowski core have been worked out by applying the Newman-Janis algorithm \cite{Newman:1965tw} and other methods \cite{Azreg-Ainou:2014pra,Azreg-Ainou:2014aqa,Azreg-Ainou:2014nra}. These algorithms also enable one to derive regular black hole solutions inspired from non-commutative geometry with a regular de-Sitter toroidal core \cite{Modesto:2010rv}. Since we will be interested in exploring the aforesaid metric in the context of astrophysical 
observations we work with the rotating counterpart in this work \cite{Ghosh:2014pba}. As expected, the rotating solution resembles the Kerr-Newman metric far from the source. When the radiating counterpart of this spacetime is considered one obtains generalization of Carmeli’s spacetime as well as Vaidya’s spacetime in suitable limits.

In this work our goal is to extract signatures of the non-linear electrodynamics charge parameter from the quasi-periodic oscillations (QPOs) observed in the power density spectrum of black holes. QPOs generally appear as peaks in the power spectrum of Low-Mass X-ray binaries (LMXRBs) which includes black hole and neutron star sources. Although rare, certain active galactic nuclei (AGNs) also exhibit QPOs in the power spectrum. It is believed that QPOs, in particular, the ones with high frequencies (henceforth referred as HFQPOs) encapsulate information regarding the nature of strong gravity associated with these compact objects and hence these can serve as effective tools to study the alternatives to \gr \cite{Bambi:2012pa,Bambi:2013fea}. The QPO frequencies are inversely proportional to the mass of the compact objects, such that for neutron stars HFQPOs $\sim$ kHz, for stellar-mass black holes they are $\sim$ hundreds of Hz while for supermassive black holes they are in the mHz order \cite{2006csxs.book.....L,vanderKlis:2000ca}. The order of magnitude of the HFQPOs can be directly attributed to the timescales associated with the motion of matter close to the compact objects. This can be directly seen from the fact that the dynamical time scale $t_d\sim r^3/GM\sim$ milli-seconds for neutron stars and stellar-mass BH sources when $r<10~r_g$ is considered \cite{2006csxs.book.....L,vanderKlis:2000ca,2008Natur.455..369G,PhysRevLett8217}. This was predicted in the 1970s which later received confirmation with the launch of NASA's Rossi X-Ray Timing Explorer satellite. Thus, QPOs provide unique opportunity to explore the nature of strong gravity and dense matter. Since we intend to investigate the nature of strong gravity, i.e. to study the role of the NED charge parameter in explaining the observed QPOs we will concentrate only on black hole sources in this work. 

The paper is organized in the following way: In \ref{S2} we discuss the black hole solution considered in this work which arises when gravity is coupled to non-linear electrodynamics. The motion of test particles in such a spacetime is studied in \ref{S3} and the dependence of the epicyclic frequencies on the background spacetime is derived. \ref{S4} is dedicated in reviewing the various theoretical models proposed to explain the observed QPOs. In \ref{S5} we compare the various model dependent QPO frequencies with observations and perform a $\chi^2$ analysis to derive the observationally favored NED charge parameter. We conclude with a summary of our results and discuss some avenues for future work in \ref{S6}.

\section{Regular rotating black holes in non-linear electrodynamics with a Minkowski core}\label{S2}
In this work we consider regular black holes in non-linear electrodynamics with an asymptotically Minkowski core. The action corresponding to non-linear electrodynamics coupled to Einstein gravity is given by \cite{Kumar:2020ltt,Ayon-Beato:2000mjt,Salazar:1987ap,Fan:2016hvf,Bronnikov:2017tnz,Toshmatov:2018cks}, 
\begin{align}
S=\int d^4 x \sqrt{-g} \bigg(\frac{\mathcal{R}}{16\pi}-\frac{L(F)}{4\pi} \bigg)
\label{Eq1}
\end{align}
where $\mathcal{R}$ is the Ricci scalar and $L(F)$ is the Lagrangian density associated with non-linear electrodynamics such that $F=F^{\mu\nu}F_{\mu\nu}/4$ is the Faraday invariant and $F_{\mu\nu}=\partial_\mu A_\nu - \partial_\nu A_\mu$ is the electromagnetic field strength tensor with $A_\mu$ the gauge field. The Maxwell theory is retrieved in the weak field limit when $L(F)=F$. Varying the action with respect to the metric yields the Einstein's equations with $L(F)$ as the source,
\begin{align}
G_{\mu\nu}=2 (L_F F_\mu^\sigma F_{\nu\sigma}-g_{\mu\nu}L(F))
\label{Eq2}
\end{align}
where $L_F=\frac{\partial L }{\partial F}$ and $G_{\mu\nu}$ is the Einstein tensor.
On the other hand, varying the action with respect to $A_\mu$ yields the equation of motion corresponding to non-linear electrodynamics,
\begin{align}
\lbrace L_F F^{\mu\nu}\rbrace_{;\mu}=0 ~~~~(^*F^{\mu\nu})_{;\nu}=0
\label{Eq3}
\end{align} 
where $^*F^{\mu\nu}=\epsilon^{\mu\nu\alpha\beta}F_{\alpha\beta}$ is the Hodge-dual of $F^{\mu\nu}$.
When the Lagrangian density corresponding to non-linear electrodynamics is considered to be,
\begin{align}
L(F)=Fe^{-\alpha(2q^2F)^{1/4}}
\label{Eq4}
\end{align}
with $\alpha=qc^2/(2G\tilde{\mathcal{M}})$ ($q$ being the charge and $\tilde{\mathcal{M}}$ the mass of the black hole), the static, spherically symmetric and asymptotically flat solution of \ref{Eq2} assumes the form,
\begin{align}
ds^2=-\Bigg(1-\frac{2\tilde{m}(r)}{r}\Bigg)dt^2 + \frac{dr^2}{\Bigg(1-\frac{2\tilde{m}(r)}{r}\Bigg)} +r^2 (d\theta^2 + sin^2\theta d\phi^2)
\label{Eq5}
\end{align}
with the mass function 
\begin{align}
\tilde{m}(r)=\tilde{\mathcal{M}}e^{-\kappa/r}
\label{1}
\end{align}
where $\kappa=\frac{q^2c^2}{2G\tilde{\mathcal{M}}}$ has dimensions of length. 
The above spacetime has two horizons \cite{Simpson:2019mud,Kumar:2020ltt}. The outer horizon corresponds to $r_+=2\tilde{\mathcal{M}} e^{W_0\big(-\frac{\kappa}{\tilde{\mathcal{M}}}\big)}$ while the inner horizon corresponds to $r_-=2\tilde{\mathcal{M}} e^{W_{-1}\big(-\frac{\kappa}{\tilde{\mathcal{M}}}\big)}$, where $W_0, W_{-1}$ corresponds to the Lambert W function.
When $r=\kappa$ the two horizons merge resulting in the extremal black hole scenario.

The above metric arises from the solution of Einstein's equations with the source,
\begin{align}
&T^0_0=-\rho(r)=\frac{-\tilde{M} \kappa}{4\pi  r^4} e^{-\kappa/{r}};\nonumber \\
&T^1_1=-\rho(r)=\frac{-\tilde{M} \kappa}{4\pi  r^4} e^{-\kappa/{r}}; \nonumber \\
&T^2_2=T^3_3=\frac{\tilde{M} \kappa}{4\pi  r^4}(1-\frac{\kappa}{2r})e^{-\kappa/r}
\label{energy}
\end{align}
The energy density and pressure described by \ref{energy} tends to zero as $r\to \infty$ and are non-singular at $r=0$. We further note that the above energy momentum tensor satisfies the weak energy condition and goes over to the Maxwell stress tensor far from the horizon \cite{Culetu:2013fsa}.

Since black holes in astrophysics are in general rotating, studying rotating black hole solutions of the above Einstein's  equations are more important. Such a solution is generated by applying the Newman Janis algorithm \cite{Newman:1965tw,Azreg-Ainou:2014pra,Azreg-Ainou:2014aqa,Azreg-Ainou:2014nra} and turns out to be  
\begin{align}
\label{metric_bardeen}
ds^{2} &=-\bigg{(} 1 - \frac{2\tilde{m}(r)r}{\tilde{\Sigma}}\bigg{)}dt^{2} - \frac{4a\tilde{m}(r)r}{\tilde{\Sigma}}\sin^{2}\theta dt d\phi + \frac{\tilde{\Sigma}}{\Delta}dr^{2} \nonumber\\
&+\tilde{\Sigma} d\theta^{2} + \bigg{(} r^{2} + a^{2} + \frac{2\tilde{m}(r)ra^{2}}{\tilde{\Sigma}}\sin^{2}\theta\bigg{)}\sin^{2}\theta d\phi^{2}
\end{align}
where, 
\begin{align}\label{metricparams_bardeen}
\tilde{\Sigma} = r^{2} + a^{2}\cos^{2}\theta ~ {,} ~ \Delta = r^{2} + a^{2} - 2\tilde{m}(r)r
\end{align}
and $\tilde{m}(r)$ is the mass function mentioned in \ref{1} such that $lim_{r\rightarrow\infty}\tilde{m}(r) = \tilde{\mathcal{M}}$ and $a$ is the spin parameter of the black hole. Since it is computationally easier to handle dimensionless quantities we scale $\kappa$ and $r$ in \ref{1} by the gravitational radius $r_g=GM/c^2$. Thus the dimensionless metric parameters correspond to the squared charge to mass ratio $k=\kappa/r_g=\frac{q^2c^4}{2G^2\tilde{\mathcal{M}}^2}$ and the spin parameter $a\equiv a/r_g$. 

The above metric goes over to the Kerr-Newman spacetime when $r>> \kappa$.
It is important to note that the above black hole solution does not have the curvature singularity at $r=0$ and has an asymptotically Minkowski core. In such a situation the energy density $\rho(r)\to 0$ as $r\to 0$ unlike a de-Sitter core where the energy density becomes constant at the core. It turns out that the curvature tensor and invariants in the above spacetime are considerably simpler than the Bardeen metric and has several physically interesting features defined by the Lambert W function \cite{Valluri:2000zz,Boonserm:2008zg,Boonserm:2010px,Boonserm:2013dua,Boonserm:2018orb,Sonoda:2013kia,Sonoda:2013jia,Culetu:2013fsa}.

In the absence of nonlinear electrodynamics when $\kappa=0$ the above black hole solution reduces to the Kerr metric.
The event horizon of the above metric is obtained by solving for the roots of $g^{rr}=\Delta=0$ which gives,
\begin{align}
r^2 + a^2 - 2r e^{-k/r} =0
\label{horizon_bardeen}
\end{align}
For the metric in \ref{metric_bardeen} to represent a black hole real, positive horizons must be present. This sets the physically allowed range of $k$ to $0\lesssim k \lesssim 0.7$ \cite{Kumar:2018ple}.

In the next section we discuss the derivation of the epicyclic frequencies of the accreting test particles in the above spacetime.

\section{Epicyclic frequencies of test particles in a stationary axisymmetric black hole spacetime}\label{S3}
This section investigates motion of massive test particles around a rotating black hole described by a stationary, axisymmetric spacetime whose metric is given by,
\begin{align}
ds^2=g_{tt}dt^2 + 2g_{t\phi}dt d\phi + g_{\phi\phi}d\phi ^2 + g_{rr}dr^2 +g_{\theta\theta} d\theta^2~,
\label{3-1}
\end{align}
We assume that the above spacetime has reflection symmetry such that $g_{\mu\nu}(r,\theta)=g_{\mu\nu}(r,-\theta)$. Stationarity and axisymmetry implies $\partial_t$ and $\partial_\phi$ are Killing vectors such that the specific energy $E$ and the specific angular momentum $L$ of the test particles are conserved. The test particles moving in such a spacetime move in an effective potential given by,
\begin{align}
V(r,\theta)=(g^{tt}-2lg^{t\phi}+l^{2}g^{\phi\phi})
\label{3-0}
\end{align}
which can be obtained from the invariance of the rest mass of the test particles such that,
\begin{align}
\dot{r}^2g_{rr}+\dot{\theta}^2g_{\theta\theta} +E^2 V(r,\theta)=-1
\label{3-2}
\end{align}
where $l=L/E$ is the impact parameter in \ref{3-0}. If we consider circular orbits in the equatorial plane the above equation reduces to
\begin{align}
E^2 V(r_0,\pi/2)=-1
\label{3}
\end{align}
where the radius of the circular orbit is denoted by $r_0$. The radial Euler-Lagrange equation leads to a quadratic equation for the angular frequency
\begin{align}
\label{3-3}
g_{tt,r}+2\Omega  g_{t\phi,r} +\Omega^{2} g_{\phi\phi,r}=0~.
\end{align}
which when solved gives 
\begin{align}
\label{3-4}
\Omega=\frac{ -g_{t\phi,r}\pm \sqrt{g_{t\phi,r}^2 - g_{tt,r}g_{\phi\phi,r}}}{ g_{\phi\phi,r}}
\end{align}
where $\pm$ sign represents prograde and retrograde orbits respectively.
We next consider slight perturbations in the motion of the test particle from the circular orbit and the equatorial plane which are denoted by,
\begin{align}
r(t)\simeq r_{\rm 0}+\delta r_{0}~e^{i\omega_r t}~;
\qquad
\theta(t)\simeq \frac{\pi}{2}+\delta\theta_{0}~e^{i\omega_\theta t}~.
\label{3-5}
\end{align}
where $\omega_r$ and $\omega_\theta$ correspond to radial and vertical epicyclic frequencies respectively. Substituting \ref{3-5} in \ref{3-2} and Taylor expanding $U(r,\theta)$ about $r=r_0$ and $\theta=\pi/2$ we get,
\begin{align}
-\delta r_{0}^2 \omega_r^2 (u^t)^2g_{rr}-\delta \theta^2 \omega_\theta^2 (u^t)^2g_{\theta\theta} + E^2 \bigg[V(r_0,\pi/2) + \frac{1}{2}\frac{\partial^2V}{\partial r^2}\Bigg|_{r=r_0,\theta=\pi/2} \delta r^2+ \frac{1}{2}\frac{\partial^2V}{\partial \theta^2}\Bigg|_{r=r_0,\theta=\pi/2}\delta \theta^2 \bigg] =-1
\label{3-6}
\end{align}
In \ref{3-6} we retain terms upto quadratic order such that the motion in the radial and the vertical direction are uncoupled. Using \ref{3} in \ref{3-6} and by equating the coefficients of $\delta r^2$ and $\delta \theta^2$ on both sides of \ref{3-6} we obtain radial and the vertical epicyclic frequencies of the test particles in the above spacetime which are respectively given by,
\begin{align}
\omega_r^2 =\frac{c^6}{G^2\tilde{M}^2} \frac{(g_{tt}+\Omega g_{t\phi})^2}{2g_{rr}}\frac{\partial^2V}{\partial r^2}\bigg|_{r=r_0,\theta=\pi/2}  \\
\omega_\theta^2 = \frac{c^6}{G^2\tilde{M}^2} \frac{(g_{tt}+\Omega g_{t\phi})^2}{2g_{\theta\theta}}\frac{\partial^2V}{\partial \theta^2}\bigg|_{r=r_0,\theta=\pi/2} 
\label{3-7}
\end{align}
In \ref{3-7} the radial and vertical epicyclic frequencies are multiplied by the factor $(c^6/G^2\tilde{M}^2)$ so that they have dimensions of frequency squared. We note that these are functions of the black hole mass $\tilde{M}$, the radius at which these oscillations are generated $r_{em}$ and the metric parameters $k$ and $a$.

\ref{Table1} below enlists some BH sources where QPOs have been discovered. We are chiefly interested in high frequency QPOs in this work which for stellar mass black holes $\sim$ hundreds of Hz while for supermassive black holes the QPO frequencies $\sim $ mHz\cite{2008Natur.455..369G,Torok:2004xs,Aschenbach:2004kj,2006csxs.book.....L,vanderKlis:2000ca,PhysRevLett8217}. This can be understood in the following way.
The deviation of the $g_{tt}$ component of the metric from the Minkowski spacetime corresponds to the Newtonian potential. In our case, if we consider the spherically symmetric metric given by \ref{Eq5} the Newtonian potential is given by,
\begin{align}
\Phi=\frac{G\tilde{\mathcal{M}}}{r}e^{-k/r}
\end{align}   
The force per unit mass can thus be evaluated, 
\begin{align}
F=-\frac{d\Phi}{dr}=\frac{G\tilde{\mathcal{M}}}{r^2}e^{-k/r}(1-k/r)=\frac{v^2}{r}
\end{align}
Therefore the velocity is given by:
\begin{align}
v\simeq\sqrt{\frac{G\tilde{\mathcal{M}}}{r}\bigg(1-\frac{k}{r}\bigg)}e^{-k/2r}
\end{align}
Taking $r\simeq 8 r_g$ and $k\simeq 0.2$ we note that 
\begin{align}
v\simeq 0.988*\sqrt{\frac{G\tilde{\mathcal{M}}}{r}0.975}
\end{align}
The dynamical timescale associated to matter rotating around the black hole is therefore given by,
\begin{align}
t_d=\frac{r}{v}\sim 1.025\sqrt{\frac{r^3}{G\tilde{\mathcal{M}}}}
\end{align}
which for a $10M_\odot$ black hole at $r\simeq 8r_g$ turns out to be $t_d \sim 1.11 ms$. 
The corresponding frequency is therefore $\sim $ hundreds of Hertz.

Theoretical models aimed to explain these QPOs are dependent on the epicyclic frequencies which are inversely proportional to the black hole mass which explains the reason for lower frequencies in supermassive BHs \cite{2008Natur.455..369G,Torok:2004xs,Aschenbach:2004kj}. 
Apart from HFQPOs, low frequency QPOs (LFQPOs) are also observed in some sources. We denote the observed HFQPOs by $\nu_{u1}$ and $\nu_{u2}$ while the low frequency QPO is denoted by $\nu_L$. 

We note that RE J1034+396 galaxy exhibits only a single QPO in its power spectrum \cite{2008Natur.455..369G,Middleton:2008fe,Jin:2020hgq,Jin:2020meg}. 
From \ref{Table1} we note that HFQPOs in the BH sources generally appear in the ratio of 3:2 and we will consider only those sources in \ref{Table1} (namely, the first five sources) which exhibit the 3:2 ratio HFQPOs. Therefore the data related to RE J1034+396 galaxy will not be used for subsequent analysis.

\begin{table}[h]
\begin{center}
\begin{tabular}{|c|c|c|c|c|}
\hline
$\rm Source $ & $\rm Mass$ & $ \nu_{u1} \pm  \Delta \nu_{u1}$ & $ \nu_{u2} \pm {\Delta} \nu_{u2}$ & $ \nu_L \pm \Delta \nu_L$\\
& $\rm (\rm M_{\rm BH ~} in ~M_\odot)$ & $(\rm in~ Hz)$ & $\rm (\rm in ~Hz)$ & $\rm (\rm in~ Hz)$\\
\hline 
$\rm GRO ~J1655-40$ & $\rm 5.4\pm 0.3$ \cite{Beer:2001cg} & $\rm 441  \rm \pm 2 $ \cite{Motta:2013wga} & $\rm 298 \rm \pm 4 $ \cite{Motta:2013wga} & $\rm 17.3 \pm \rm 0.1 $ \cite{Motta:2013wga}\\ 
\hline
$\rm XTE ~J1550-564$ & $\rm 9.1\pm 0.61$ \cite{Orosz:2011ki} & $\rm 276 \rm \pm 3 $ & $\rm 184  \pm 5 $ & $ -$\\
\hline
$\rm GRS ~1915+105$ & $\rm 12.4^{+2.0}_{-1.8}$ \cite{Reid:2014ywa} & $\rm 168  \pm 3 $ & $\rm 113  \pm 5 $ & $\rm - $\\
\hline
$\rm H ~1743+322$ & $\rm 8.0-14.07$ \cite{Pei:2016kka,Bhattacharjee:2019vyy,Petri:2008jc} & $\rm 242 \pm 3 $ & $\rm 166  \pm 5 $ & $\rm - $\\
\hline
$\rm Sgr~A^*$ & $\rm (3.5-4.9)$ & $\rm (1.445 \pm 0.16)$ & $\rm (0.886 \pm 0.04)$ & $ - $\\
 & $\rm ~\times 10^6$ \cite{Ghez:2008ms,Gillessen:2008qv} & $\rm ~\times 10^{-3} $ \cite{Torok:2004xs,Stuchlik:2008fy} & $\rm ~\times 10^{-3} $ \cite{Torok:2004xs,Stuchlik:2008fy} & $ - $\\
 \hline
 $\rm RE J1034+396$ & $\rm (1-4) ~\times 10^6$ & $\rm (2.5-2.8) \rm ~\times 10^{-4}$ & $-$ & $ - $\\
 &  \cite{2008Natur.455..369G,2012MNRAS.420.1825J,Czerny:2016ajj,Chaudhury:2018jzz} &  \cite{2008Natur.455..369G,Middleton:2008fe,Jin:2020hgq,Jin:2020meg} &   & \\
\hline
\end{tabular}
\caption{Black hole sources where high frequency QPOs (HFQPOs) are observed}
\label{Table1}
\end{center}

\end{table}




\begin{table}[t!]
\begin{center}
\begin{tabular}{|c|c|c|c|}
\hline
$\rm Model $ & $ \nu_1 $ & $ \nu_2$ &  $\nu_3 $ \\
\hline 
$\rm Relativistic ~Precession ~Model ~(kinematic)$ \cite{Stella:1997tc,PhysRevLett8217,Stella_1999} & $\rm \nu_\phi$ & $\rm \nu_\phi-\nu_r $ & $\rm \nu_\phi-\nu_\theta$ \\ 
\hline
$\rm Tidal ~Disruption~Model~(kinematic)$ \cite{Cadez:2008iv,Kostic:2009hp,Germana:2009ce} & $\rm \nu_\phi + \nu_r$ & $\rm \nu_\theta $ & $-$ \\
\hline
$\rm Parametric ~Resonance~Model~(resonance)$ \cite{Kluzniak:2002bb,Abramowicz:2003xy,Rebusco:2004ba} & $\rm \nu_\theta$ & $\rm \nu_r$ & $-$\\ \hline
$\rm Forced ~Resonance~Model~1 ~(resonance)$ \cite{Kluzniak:2002bb} & $\rm \nu_\theta$ & $\rm \nu_\theta-\nu_r$ & $-$\\ \hline
$\rm Forced ~Resonance~Model~2 ~(resonance)$ \cite{Kluzniak:2002bb} & $\rm \nu_\theta+ \nu_r$ & $\rm \nu_\theta$ & $-$\\ \hline
$\rm Keplerian ~Resonance~Model~1 ~(resonance)$ \cite{Nowak:1996hg} & $\rm \nu_\phi$ & $\rm \nu_r$ & $-$\\ \hline
$\rm Keplerian ~Resonance~Model~2 ~(resonance)$ \cite{Nowak:1996hg} & $\rm \nu_\phi$ & $\rm 2\nu_r$ & $-$\\ \hline
$\rm Keplerian ~Resonance~Model~3 ~(resonance)$ \cite{Nowak:1996hg} & $\rm 3 \nu_r$ & $\rm \nu_\phi$ & $-$\\ \hline
$\rm Warped ~Disk~Oscillation~Model~ ~(resonance)$ & $\rm 2\nu_\phi-\nu_r$ & $\rm 2(\nu_\phi-\nu_r)$ & $-$\\ \hline
$\rm Non-axisymmetric ~Disk~Oscillation~Model~1 ~(resonance)$ & $\rm \nu_\theta$ & $\rm \nu_\phi-\nu_r$ & $-$\\ \hline
$\rm Non-axisymmetric ~Disk~Oscillation~Model~2 ~(resonance)$  \cite{Torok:2010rk,Torok:2011qy,Kotrlova:2020pqy} & $\rm 2\nu_\phi-\nu_\theta$ & $\rm \nu_\phi-\nu_r$ & $-$\\ \hline
\end{tabular}
\caption{Theoretical models explaining the HFQPOs in black holes.}
\label{Table2}
\end{center}
\end{table}
\section{Theoretical Models explaining quasi-periodic oscillations in the black hole power spectrum}\label{S4}
This section is dedicated in reviewing some of the existing theoretical models proposed to explain the observed HFQPOs in BHs \cite{Stella:1997tc,PhysRevLett8217,Stella_1999,Cadez:2008iv,Kostic:2009hp,Germana:2009ce,Kluzniak:2002bb,Abramowicz:2003xy,Rebusco:2004ba,Nowak:1996hg,Torok:2010rk,Torok:2011qy,Kotrlova:2020pqy,1980PASJ...32..377K,Perez:1996ti,Silbergleit:2000ck,Dexter:2013sxa,Rezzolla:2003zy,Rezzolla:2003zx}. These models mainly aim to explain the commensurability of the QPO frequencies. \ref{Table2} presents the mathematical expressions for the model dependent QPO frequencies where the upper and lower HFQPOs are denoted by $\nu_1$ and $\nu_2$ while the low frequency QPO is denoted by $\nu_3$. From the table it is evident that the theoretical QPO frequencies are linear combinations of the angular frequency, the radial and the vertical epicyclic frequencies respectively given by $\nu_\phi=\frac{\omega_\phi}{2\pi}$, $\nu_r=\frac{\omega_r}{2\pi}$ and $\nu_\theta=\frac{\omega_\theta}{2\pi}$.
Since the theoretical QPO frequencies depend purely on the background metric and not on the complex accretion physics, QPOs can potentially extract more accurate information about the background spacetime compared to other available observations, e.g., the iron line or the continuum-fitting methods. 

\begin{itemize}
\item {\bf Parametric resonance model: } The 3:2 ratio of the observed twin-peak HFQPOs (with frequencies $\rm \nu_{u1}$ and $\rm \nu_{u2}$) in BH and NS sources indicates that QPOs might be a consequence of resonance between various oscillation modes in the accretion disk \cite{2001A&A...374L..19A,Kluzniak:2002bb,2001AcPPB..32.3605K}. When we considered circular, equatorial motion of test particles and slight perturbations $\delta r$ and $\delta \theta$ in the orbit we tacitly assumed that the perturbations obey the equations of simple harmonic motion such that 
\begin{align}
\delta \ddot{ r}+\omega_r^2 \delta r=0~; \qquad \delta \ddot{\theta}+\omega_\theta^2 \delta \theta=0~. 
\label{S4-1}
\end{align} 
with frequencies $\omega_{r}=2\pi \nu_r$ and $\omega_{\theta}=2\pi \nu_\theta$ respectively. This implies that the motion in the radial and the vertical direction are uncoupled which is applicable to both thin disks as well as to more general accretion flow models, e.g. accretion tori \cite{2001A&A...374L..19A,Kluzniak:2002bb}
The presence of dissipation and pressure effects necessitate including forcing terms in \ref{S4-1} 
\begin{align}
\delta \ddot{r}+\omega_r^2 \delta r=\omega_r^{2}F_{r}(\delta r,\delta \theta, \delta\dot{r}, \delta\dot{\theta})~; 
\qquad 
\delta \ddot{\theta}+\omega_\theta^2 \delta \theta=\omega_{\theta}^{2} F_{\theta}(\delta r,\delta \theta, \delta\dot{r}, \delta\dot{\theta})~.
\label{S4-2}
\end{align}
which are in general some non-linear functions of their arguments. The forms of $F_r$ and $F_\theta$ are determined by the accretion flow model \cite{Abramowicz:2003xy, Horak:2004hm}. 

In the parametric resonance model it is assumed that the radial epicyclic motion triggers the vertical epicyclic motion since random fluctuations in thin disks are expected to have $\delta r \gg \delta \theta$ \cite{Kluzniak:2002bb,2001A&A...374L..19A,Abramowicz:2003xy,2005A&A...436....1T,Rebusco:2004ba}. In this scenario \ref{S4-2} assume the form,
\begin{align}
\delta \ddot{ r}+\omega_r^2 \delta r=0 ~~~~~~\delta \ddot{\theta}+\omega_\theta^2 \delta \theta= -\omega_\theta^2 \delta r \delta \theta
\label{S4-3}
\end{align}
such that $\delta r=B \cos (\omega_r t)$ (where $B$ is a constant) and the equation for $\delta\theta$ assumes the form of the Matthieu equation \cite{1969mech.book.....L} given by
\begin{align}
\delta \ddot{\theta}+\omega_\theta^2 (1+B\delta r)\delta \theta =0 
\label{S4-4}
\end{align}
and is excited when \cite{Rebusco:2004ba,1969mech.book.....L,Abramowicz:2003xy}
\begin{align}
\frac{\nu_r}{\nu_\theta}=\frac{2}{n}~, 
\qquad
\rm where ~ n \in ~positive ~integers
\label{S4-5}
\end{align} 
For rotating non-singular BHs $\nu_\theta > \nu_r$ \cite{Stuchlik:2008fy} (where $\nu_\theta=\omega_\theta/2\pi$ and $\nu_r=\omega_r/2\pi$) and therefore $n=3$ gives rise to the strongest resonance which naturally explains the observed $3:2$ ratio of the HFQPOs. This result has been confirmed by analytical calculations and numerical simulations \cite{Rebusco:2004ba,Horak:2004hm,Abramowicz:2003xy}.

\item {\bf Forced Resonance Model:} Thin disks or nearly Keplerian disks are more likely to exhibit parametric resonance \cite{2001A&A...374L..19A,Kluzniak:2002bb,2001AcPPB..32.3605K}. In a more realistic flow however, couplings between $\delta r$ and $\delta\theta$ are expected in addition to parametric resonance, arising from pressure, viscous or magnetic stresses present in the accretion flow leading to non-zero forcing terms \cite{2005A&A...436....1T}. Numerical simulations confirm that pressure couplings often 
lead to a resonant forcing of vertical oscillations by radial oscillations \cite{2001A&A...374L..19A,2004ApJ...603L..93L}. Since the physics of accretion that leads to such resonant couplings is not very clearly understood the forcing terms are given by some mathematical ansatz, e.g.,
\begin{align}
\delta \ddot{\theta}+\omega_\theta^2 \delta \theta =-\omega_\theta^2\delta r \delta \theta + \mathcal{F_\theta}(\delta\theta) 
\label{S4-7}
\end{align}
such that $\delta r=B cos(\omega_r t)$ while $\mathcal{F_\theta}$ corresponds to the non-linear terms in $\delta\theta$. 
It can be shown that \ref{S4-7} has solutions of the form,
\begin{align}
\frac{\nu_\theta}{\nu_r}=\frac{m}{n} \rm~~~~~where ~m ~and~ n ~are~ natural~ numbers
\label{S4-8}
\end{align}
The presence of non-linear couplings lead to forced resonances, e.g. $m:n=3:1$ and $m:n=2:1$, apart from the $3:2$ parametric resonance
which permits resonance between combinations of frequencies, e.g. $\nu_\theta-\nu_r$, $\nu_\theta+\nu_r$.
The forced resonance model comprises of $3:1$ and $2:1$ forced resonances. In $3:1$ forced resonance model (denoted by Forced Resonance Model 1 or FRM1), the lower HFQPO is given by $\nu_{2}=f_-= \nu_\theta-\nu_r$ while the upper HFQPO is given by $\nu_{1}=\nu_\theta$. For 
$2:1$ forced resonance model the upper and lower high frequency QPOs are denoted by $\nu_{1}=f_+= \nu_\theta+\nu_r$ while $\nu_{2}=\nu_\theta$ respectively.

\item {\bf Keplerian resonance model: } In Keplerian resonance model one considers resonance between the orbital motion and the radial epicyclic motion \cite{2005A&A...436....1T,2001PASJ...53L..37K,2001A&A...374L..19A, Nowak:1996hg}. Keplerian resonance might arise under two possible circumstances: (a) trapping of non-axisymmetric g-mode oscillations induced by a corotation resonance in the inner region of relativistic thin accretion disks \cite{2001PASJ...53L..37K} and (b) when a pair of spatially separated coherent vortices with opposite vorticities oscillating with radial epicyclic frequencies couple with the spatially varying orbital angular frequencies \cite{2005A&A...436....1T,2010tbha.book.....A}. However, it was realized that g-mode oscillations are dampened by corotation resonance \cite{Li:2002yi,2003PASJ...55..257K} such that models invoking Keplerian resonance may not be very useful in explaining the HFQPOs in microquasars. Keplerian model consists of resonance between (i) $\nu_1=\nu_\phi$ and $\nu_2=\nu_r$ (which we denote as Keplerian Resonance Model 1 or KRM1), (ii) $\nu_1=\nu_\phi$ and $\nu_2=2\nu_r$ (which we denote as Keplerian Resonance Model 2 or KRM2) and (iii) $\nu_1=3\nu_r$  and $\nu_2=\nu_\phi$ (which we denote as Keplerian Resonance Model 3 or KRM3).

\item {\bf Warped disk oscillation model:} This model which assumes a somewhat unusual disk geometry \cite{Torok:2011qy,Yagi:2016jml} explains the HFQPOs in terms of non-linear resonances between the relativistic disk deformed by a warp with various disk oscillation modes \cite{2001PASJ...53....1K,2004PASJ...56..559K,2004PASJ...56..905K,2005PASJ...57..699K,2008PASJ...60..111K}. Such resonances comprise of horizontal resonances inducing g-mode and p-mode oscillations as well as vertical resonances which can induce only the g-mode oscillations \cite{2004PASJ...56..559K}. The origin of such resonances can be attributed to the non-monotonic variation of the radial epicyclic frequency with the radial distance $r$ \cite{2004PASJ...56..559K}. According to this model the upper high frequency QPO is given by $\nu_{1}=2\nu_\phi -\nu_r$ while the lower HFQPO is given by $\nu_{2}=2(\nu_\phi -\nu_r)$.

\item {\bf Non-axisymmetric disk oscillation model:} Non-axisymmetric Disk-Oscillation models consider various combinations of non-axisymmetric disc oscillation modes as the origin of the HFQPOs \cite{2004ApJ...617L..45B,2005AN....326..849B,2005ragt.meet...39B,Torok:2011qy,Kotrlova:2020pqy}. These models which are variants of the Relativistic Precession model \cite{Torok:2011qy} include non-geodesic effects in the accretion flow by modelling the flow in terms of a slightly non-slender pressure-supported perfect fluid torus \cite{Torok:2015tpu,Sramkova:2015bha,Kotrlova:2020pqy}. Two variants of non-axisymmetric disk-oscillation models are proposed in the literature: (i) the first (denoted by NADO1) \cite{2004ApJ...617L..45B,2005AN....326..849B,2005ragt.meet...39B}) assumes resonance between the vertical epicyclic frequency ($\nu_1=\nu_\theta$) with $m=-1$ non-axisymmetric radial epicyclic frequency ($\nu_2=\nu_\phi-\nu_r$) while the second (denoted by NADO2) considers resonance between the $m=-1$ non-axisymmetric radial epicyclic frequency ($\nu_2=\nu_\phi-\nu_r$) with the $m=-2$ non-axisymmetric vertical epicyclic frequency ($\nu_1=2\nu_\phi-\nu_\theta$) \cite{Torok:2010rk,Torok:2011qy,Kotrlova:2020pqy} where $m$ refers to the azimuthal wave number of the non-axisymmetric perturbation. However, the physical mechanisms inducing couplings between an axisymmetric and a non-axisymmetric mode or between the pairs non-axisymmetric modes are yet not very well-understood \cite{Horak:2008zg,Torok:2011qy}.

\item {\bf Relativistic precession model:} Models discussed so far are examples of resonance models.
Relativistic precession model \cite{Stella:1997tc,PhysRevLett8217} is a kinematic model which considers local motion of plasma in the accretion disk as the origin of QPOs. This model was initially proposed to address the HFQPOs in neutron star sources and then subsequently extended for black holes \cite{Stella_1999}. In this model the upper and lower high frequency QPOs are associated with the orbital anugular frequency $\nu_\phi$ and the periastron precession frequency $\nu_\phi-\nu_r$. This model also addresses the observed low frequency QPO in black holes in terms of the nodal precession frequency $\nu_\phi-\nu_\theta$.

\item {\bf Tidal disruption model:} Another example of a kinematic model is the Tidal Disruption Model \cite{Cadez:2008iv,Kostic:2009hp,Germana:2009ce} which aims to explain the observed HFQPOs in terms of plasma orbiting the central object which may get tidally stretched forming ring-like features along the orbit giving rise to the modulation in the observed flux in the black hole power spectrum. According to this model $\nu_1=\nu_\phi+\nu_r$ and $\nu_2=\nu_\phi$.

\end{itemize}

We note from the above discussion that the theoretical QPO frequencies depend on the metric parameters $a$, $k$, the black hole mass $\tilde{M}$ and the radius at which the QPO frequencies are generated $r_{em}$. Therefore we can directly constrain $k$ and not $q$ from the observations. Once we know $k$ we can calculate $q$ by using the previously determined masses. This is an artifact of the dependence of the Lagrangian density on mass. In this work $k$ is denoted to be the non-linear electrodynamics (NED) charge parameter.

The QPO models we have considered here assume that the resonances involved in giving rise to the QPOs in the power spectrum are generated at the same circular orbit given by $r_{\rm em}$ \cite{2016A&A...586A.130S,Yagi:2016jml,Kotrlova:2020pqy}. This assumption holds true for the kinematic and the resonant models which we have discussed above. Apart from these there also exist certain diskoseismic models which are based on the assumption that the oscillatory modes giving rise to the HFQPOs are emitted at different radii of the accretion disk \cite{1980PASJ...32..377K,Perez:1996ti,Silbergleit:2000ck}. Magnetohydrodynamic simulations \cite{Tsang:2008fz,Fu:2008iw,Fu:2010tf} reveal that such models cannot explain the 3:2 HFQPOs adequately and therefore those models are not considered in this work. Further, we do not aim to extract the mass of the black hole sources from the observed QPO frequencies but rather use the previously estimated masses obtained from other independent observations, e.g., optical/NIR photometry (see \ref{Table1}).

\section{Estimating the magnetic monopole charge from the observed QPOs}\label{S5}
In this section we aim to derive the most favored value of $k$ from observations related to QPOs. In order to accomplish this we compare the observed QPO frequencies given in \ref{Table1} with the model dependent QPO frequencies in \ref{Table2} and calculate the joint-$\chi^2$ which is given by,
\begin{align}
\label{S5-1}
\chi ^2 (k)=\sum_{j=1}^5 \frac{\lbrace \nu_{\textrm{u1}{,j}}-\nu_1(k,a_{\rm min},\tilde{M}_{\rm min},r_{\rm em, min}) \rbrace ^2}{\sigma_{\nu_{\rm u1},j}^2}  
+ \sum_{j=1}^5 \frac{\lbrace \nu_{\textrm{u2}{,j}}-\nu_2(k,a_{\rm min},\tilde{M}_{\rm min},r_{\rm em, min}) \rbrace ^2}{\sigma_{\nu_{\rm u2}, j}^2}~,
\end{align}
where $\nu_1$ and $\nu_2$ are the model dependent QPO frequencies, $\nu_{u1}$ and $\nu_{u2}$ are the observed QPO frequencies with errors given by $\sigma^2_{\nu_{\rm u1}}$ and $\sigma^2_{\nu_{\rm u2}}$ respectively. From \ref{Table2} it is clear that the theoretical QPO frequencies depend on the orbital frequency and the epicyclic frequencies which are functions of the metric parameters $k$, $a$, $\tilde{M}$ and the emission radius $r_{\rm em}$. Since our goal is to determine the most favored value of $k$ we do not minimize the joint-$\chi^2$ with respect to all the four parameters. Instead we divide them into two categories \cite{1976ApJ...210..642A} namely, (a) the ``interesting parameters" (which is $k$) and (b) the ``uninteresting parameters" (here $a$, $\tilde{M}$ and $r_{\rm em}$), derived from $\chi^2$ minimization for various choices of the ``interesting parameters". We note that we do not estimate the black hole mass from the present analysis but use masses of these sources independently obtained from optical/NIR photometry (see \ref{Table1}).

\begin{table}[t!]
\begin{center}
\hspace*{-2cm}
\begin{tabular}{|p{1.8cm}|p{2.8cm}|p{2.8cm}|p{3cm}|p{3cm}| p{3.5cm}| }
\hline

$\rm Comparison$ & $\rm GRO ~J1655-40  $ & $\rm XTE ~J1550-564 $ & $\rm GRS ~1915+105 $ & $\rm H ~1743+322 $ & $\rm Sgr~A^*$\\
$\rm of ~mass $ & $\rm   $ & $\rm  $ & $\rm  $ & $\rm  $ & $\rm $\\
$\rm estimates$ & $\rm   $ & $\rm  $ & $\rm  $ & $\rm  $ & $\rm $\\
$\rm (\rm in ~M_\odot) $ & $\rm   $ & $\rm  $ & $\rm  $ & $\rm  $ & $\rm $\\
\hline 
$\rm $ & $\rm $  & $\rm $ & $\rm  $ & $ \rm $ & $\rm $\\
$\rm Previous $ & $\rm 5.4\pm 0.3$ \cite{Beer:2001cg} & $ \rm 9.1\pm 0.61$ \cite{Orosz:2011ki}  & $ \rm 12.4^{+2.0}_{-1.8} $ \cite{Reid:2014ywa} & $ \rm 8.0-14.07$   & $ \rm (3.5-4.9) \rm \times 10^{-3}$\\ 
$\rm constraints$ & $ $ & $ $ & $ $ & $ $ \cite{Pei:2016kka,Bhattacharjee:2019vyy,Petri:2008jc}  & $  $ \cite{Ghez:2008ms,Gillessen:2008qv} \\

\hline
$\rm $ & $\rm $  & $\rm $ & $\rm  $ & $ \rm $ & $\rm $\\
$\rm RPM$ & $\rm 5.13~(k\sim 0)$  & $\rm 9.3~ (k\sim 0)$ & $\rm 13.99~ (k\sim 0)$ & $ \rm 12.06~(k\sim 0)$ & $\rm 4.0\times 10^6~(k\sim 0)$\\

\hline
$\rm $ & $\rm $  & $\rm $ & $\rm  $ & $ \rm $ & $\rm $\\
$\rm TDM$ & $\rm 7.0 ~ (k\sim 0.1)$  & $\rm 6.8 ~ (k\sim 0.1)$ & $\rm  8.6~( k\sim 0.1)$ & $ \rm  8.3~(k\sim 0.1 )$ & $\rm  4.4\times 10^6~(k\sim 0.1)$\\
\hline 
$\rm $ & $ $  & $ $ & $\rm  $ & $ \rm $ & $\rm $\\
$\rm PRM$ & $ 5.1\rm ~( k\sim 0.1)$  & $8.49 \rm ~(k\sim 0.1)$ & $\rm  13.5~(k\sim 0.1)$ & $ \rm 10.98~(k\sim 0.1)$ & $\rm 3.5\times 10^6~ (k\sim 0.1)$\\
\hline
$\rm $ & $\rm $  & $\rm $ & $\rm  $ & $ \rm $ & $\rm $\\
$\rm FRM1$ & $5.1 \rm ~(k\sim 0.4)$  & $\rm  9.61~ (k\sim 0.4)$ & $\rm 11.5~ (k\sim 0.4)$ & $ \rm  9.37~(k\sim 0.4)$ & $\rm  3.7\times 10^6 ~(k\sim 0.4)$\\
 \hline
 $\rm $ & $\rm $  & $\rm $ & $\rm  $ & $ \rm $ & $\rm $\\
 $\rm FRM2$ & $\rm 7.1 ~(k\sim 0.3)$  & $\rm  7.29~(k\sim 0.3)$ & $\rm  8.45~(k\sim 0.3)$ & $ \rm 7.75~(k\sim 0.3)$ & $\rm  3.66\times 10^6~(k\sim 0.3)$\\
 \hline
$\rm $ & $\rm $  & $\rm $ & $\rm  $ & $ \rm $ & $\rm $\\
$\rm KRM1$ & $\rm 5.1 (k\sim 0.1)$  & $\rm  8.49 (k\sim 0.1)$ & $\rm  12.7 (k\sim 0.1)$ & $ \rm  8.1 (k\sim 0.1)$ & $\rm 3.5 \times 10^6(k\sim 0.1)$\\
\hline 
$\rm $ & $\rm $  & $\rm $ & $\rm  $ & $ \rm $ & $\rm $\\
$\rm KRM2$ & $\rm  5.36~(k\sim 0)$  & $\rm  8.94~(k\sim 0)$ & $\rm  14.1~ (k\sim 0)$ & $ \rm  13.08~(k\sim 0)$ & $\rm  4.8\times 10^6~(k\sim 0)$\\
\hline 
$\rm $ & $\rm $  & $\rm $ & $\rm  $ & $ \rm $ & $\rm $\\
$\rm KRM3$ & $ 5.67\rm ~(k\sim 0)$  & $\rm  9.65~(k\sim 0)$ & $\rm 12.71~ (k\sim 0)$ & $ \rm 10.33~(k\sim 0)$ & $\rm  4.2\times 10^6~(k\sim 0)$\\
\hline
$\rm $ & $\rm $  & $\rm $ & $\rm  $ & $ \rm $ & $\rm $\\ 
$\rm WDOM$ & $\rm  5.62~(k\sim 0.2)$  & $\rm  9.45~(k\sim 0.2)$ & $\rm 11.89~ (k\sim 0.2)$ & $ \rm  11.43~(k\sim 0.2)$ & $\rm  3.5\times 10^6~ (k\sim 0.2)$\\
\hline 
$\rm $ & $\rm $  & $\rm $ & $\rm  $ & $ \rm $ & $\rm $\\
$\rm NADO1$ & $\rm 5.54~(k\sim 0.4)$  & $\rm 9.42 ~(k\sim 0.4)$ & $\rm  11.99~(k\sim 0.4)$ & $ \rm  9.55~(k\sim 0.4)$ & $\rm 3.5\times 10^6~ (k\sim 0.4)$\\
\hline
$\rm $ & $\rm $  & $\rm $ & $\rm  $ & $ \rm $ & $\rm $\\
$\rm NADO2$ & $\rm 5.12~ (k\sim 0.3)$  & $\rm   9.14~(k\sim 0.3)$ & $\rm   13.39~(k\sim 0.3)$ & $ \rm 12.26~ (k\sim 0.3)$ & $\rm  4.9\times 10^6~(k\sim 0.3)$\\
\hline\hline

\end{tabular}
\caption{The above table presents the mass estimates of the BH sources considered in \ref{Table1} from $\chi^2$ minimization. The previously estimated masses are also reported.
}
\label{Table3}
\end{center}

\end{table}

\begin{figure}[t!]
\centering
\includegraphics[scale=0.58]{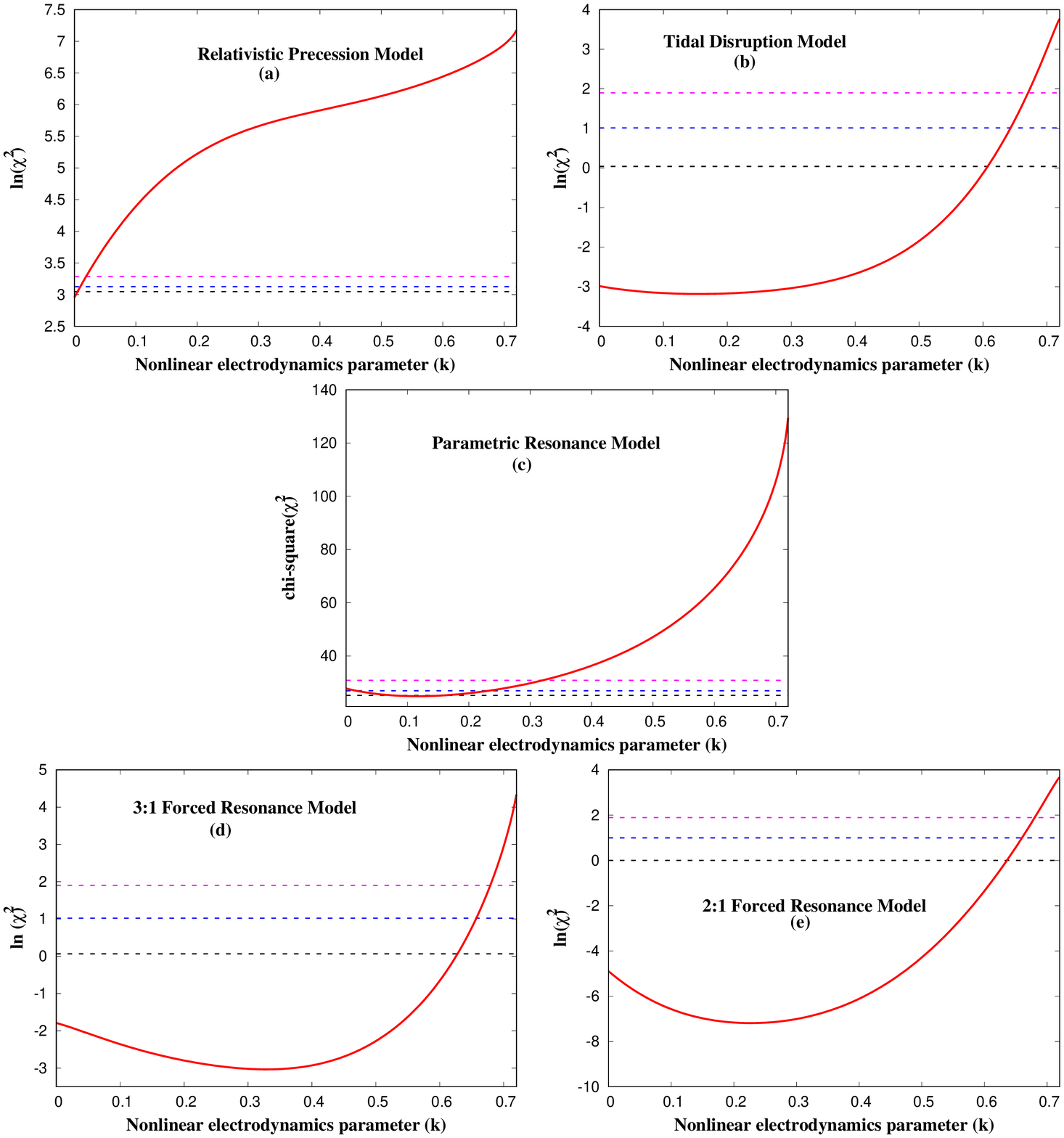}
\caption{The above figure demonstrates the variation of $\chi^2$ with the non-linear electrodynamics charge parameter $k$ assuming the following QPO models: (a) Relativistic Precession Model (RPM), (b) Tidal Disruption Model (TDM), (c) Parametric Resonance Model (PRM), (d) 3:1 Forced Resonance Model (FRM1) and (e) 2:1 Forced Resonance Model (FRM2).  The black, blue and magenta lines denote respectively the 68\%, 90\% and 99.7\% confidence intervals corresponding to $\Delta\chi^2=1$, $\Delta\chi^2=2.71$ and $\Delta\chi^2=6.63$ from $\chi^2_{min}$.} 
\label{Fig_07}
\end{figure}

\begin{figure}[t!]
\centering
\includegraphics[scale=0.58]{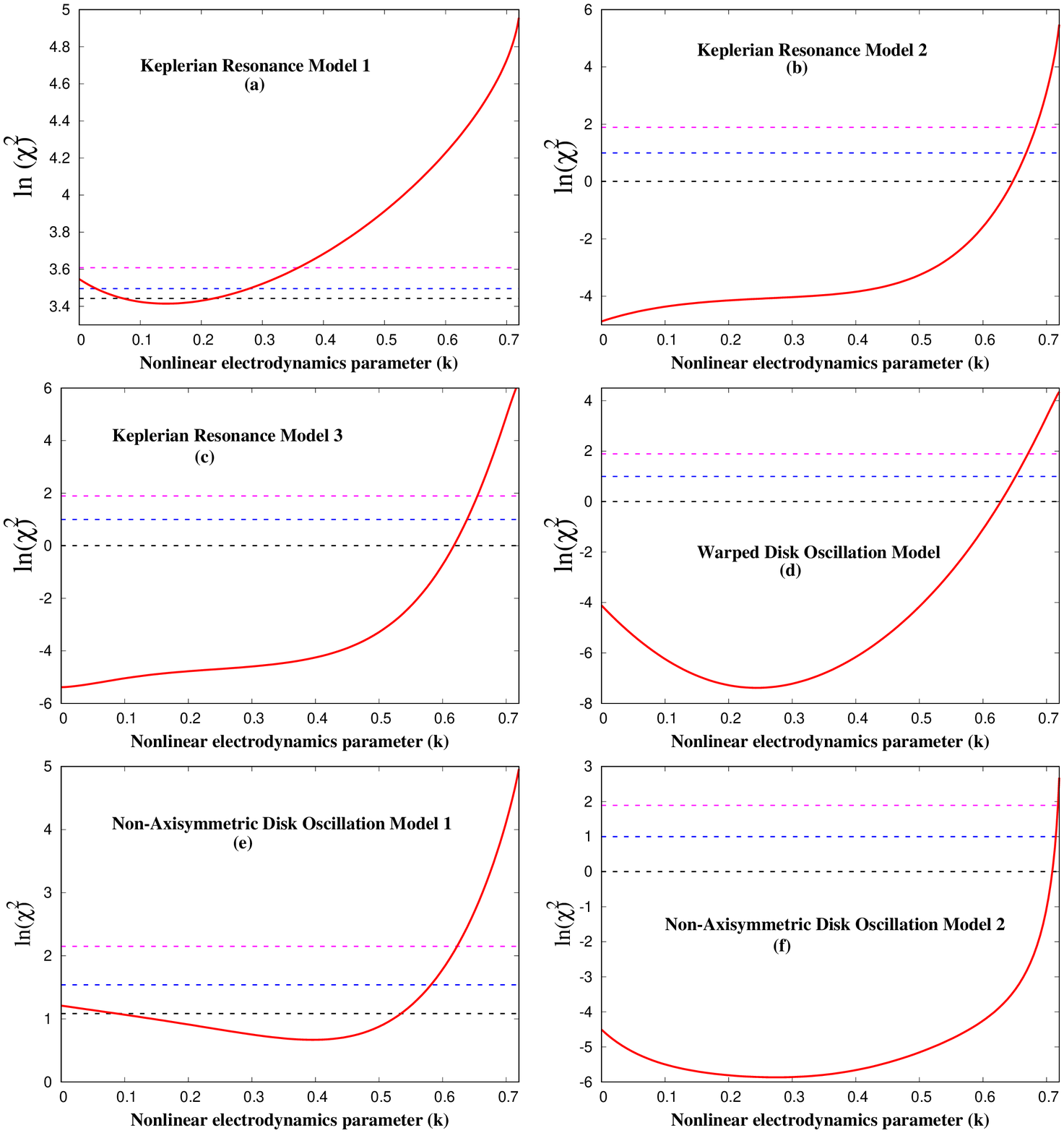}
\caption{The above figure demonstrates the variation of $\chi^2$ with the non-linear electrodynamics charge parameter $k$ assuming the following models: (a) Keplerian Resonance Model 1, (b) Keplerian Resonance Model 2, (c) Keplerian Resonance Model 3, (d) Warped Disc Oscillation Model, (e) Non-axisymmetric Disc Oscillation Model 1 and (f) Non-axisymmetric Disc Oscillation Model. The black, blue and magenta lines denote respectively the 68\%, 90\% and 99.7\% confidence intervals corresponding to $\Delta\chi^2=1$, $\Delta\chi^2=2.71$ and $\Delta\chi^2=6.63$ from $\chi^2_{min}$.
}
\label{Fig_08}
\end{figure}

\begin{table}[t!]
\vskip-1.8cm
\begin{center}
\hspace*{-2cm}
\begin{tabular}{|p{1.8cm} |p{3cm}|p{3.5cm}|p{3cm}|p{2.8cm}| p{3.3cm}| }
\hline

$\rm Comparison$ & $\rm GRO ~J1655-40  $ & $\rm XTE ~J1550-564 $ & $\rm GRS ~1915+105 $ & $\rm H ~1743+322 $ & $\rm Sgr~A^*$\\
$\rm of ~spin $ & $\rm   $ & $\rm  $ & $\rm  $ & $\rm  $ & $\rm $\\
$ \rm estimates $ & $\rm   $ & $\rm  $ & $\rm  $ & $\rm  $ & $\rm $\\
$ \rm  $ & $\rm   $ & $\rm  $ & $\rm  $ & $\rm  $ & $\rm $\\
\hline 
$\rm Previous$ & $\rm a\sim 0.65-0.75 $ \cite{Shafee_2005} & $ \rm -0.11<a<0.71$ \cite{Steiner:2010bt}  & $ a\sim \rm 0.98 $ \cite{McClintock:2006xd} & $ a=\rm 0.2 \pm 0.3 $ \cite{Steiner:2011kd} & $ \rm a\sim 0.9$ \cite{Moscibrodzka:2009gw}\\ 
$\rm constraints$ & $ \rm a\sim 0.94-0.98$ \cite{Miller:2009cw} &  $ $ & $ a\sim \rm 0.7 $ \cite{2006MNRAS.373.1004M} & $ \rm $   & $ \rm a\sim 0.5 $ \cite{Shcherbakov:2010ki} \\
$\rm $ & $ \rm a=0.29\pm 0.003  $ \cite{Motta:2013wga} & $ \rm a= 0.55^{+0.15}_{-0.22}$ \cite{Steiner:2010bt}  & $ \rm a\sim 0.6-0.98 $ \cite{Blum:2009ez} &   $   $   & $ \rm a=0.9959\pm 0.0005$ \cite{2010MmSAI..81..319A}\\ 
$\rm $ & $ $  & $ $ & $ \rm a\sim 0.4-0.98$ \cite{Mills:2021dxs} &  $  $   & $ \rm a \lesssim  0.1 $ \cite{Fragione:2020khu} \\
\hline
$\rm $ & $\rm $  & $\rm $ & $\rm   $ & $ \rm $ & $\rm $\\ 
$\rm RPM$ & $\rm 0.3~(k\sim 0)$  & $\rm 0.4~(k\sim 0)$ & $\rm 0.3~ (k\sim 0)$ & $ \rm  0.5~(k\sim 0)$ & $\rm 0.97~ (k\sim 0)$\\
\hline
$\rm $ & $\rm $  & $\rm  $ & $\rm  $ & $ \rm$ & $\rm $\\
$\rm TDM$ & $\rm 0.1 ~(k\sim 0.1)$  & $\rm 0.2 ~(k\sim 0.1)$ & $\rm -0.3 ~ (k\sim 0.1)$ & $ \rm -0.3 ~(k\sim 0.1)$ & $\rm 0.9~ (k\sim 0.1)$\\
\hline
$\rm $ & $\rm $  & $\rm  $ & $\rm  $ & $\rm  $ & $\rm  $ \\
$\rm PRM$ & $\rm 0.8~(k\sim 0.1)$  & $\rm 0.8~(k\sim 0.1)$ & $\rm 0.8~ (k\sim 0.1)$ & $ 0.89\rm  ~(k\sim 0.1)$ & $\rm ~ 0.9(k\sim 0.1)$\\
\hline
$\rm $ & $\rm $ & $ $ & $\rm $ & $ \rm $   &  $\rm   $\\
$\rm FRM1$ & $\rm -0.1~(k\sim 0.4)$  & $\rm 0.1~(k\sim 0.4)$ & $\rm -0.3~ (k\sim 0.4)$ & $ -0.1\rm  ~(k\sim 0.4)$ & $\rm 0.58~ (k\sim 0.4)$\\
\hline
 $\rm $ & $\rm  $ & $\rm  $ & $ \rm $ & $ $ &  $ $ \\
$\rm FRM2$ & $  -0.1\rm ~(k\sim 0.3)$  & $\rm -0.1~(k\sim 0.3)$ & $\rm ~ -0.4(k\sim 0.3)$ & $ \rm -0.3 ~(k\sim 0.3)$ & $\rm 0.695 ~ (k\sim 0.3)$\\
 \hline
$\rm $ & $\rm  $  & $\rm  $ & $\rm  $ & $\rm  $  &  $ $ \\
$\rm KRM1$ & $\rm 0.8997~(k\sim 0.1)$  & $\rm 0.8997~(k\sim 0.1)$ & $\rm 0.899~ (k\sim 0.1)$ & $ \rm  0.89~(k\sim 0.1)$ & $\rm 0.8997~ (k\sim 0.1)$\\
\hline 
$\rm $ & $\rm  $  & $\rm  $ & $\rm  $ & $\rm  $  &  $ $ \\
$\rm KRM2$ & $\rm 0.32~(k\sim 0)$  & $\rm 0.36~(k\sim 0)$ & $\rm 0.32~ (k\sim 0)$ & $ \rm  0.6~(k\sim 0)$ & $\rm 0.97~ (k\sim 0)$\\
\hline 
$\rm $ & $\rm  $  & $\rm $ & $\rm  $ & $\rm  $  &  $ $\\
$\rm KRM3$ & $\rm 0.22~(k\sim 0)$  & $\rm 0.3~(k\sim 0)$ & $\rm 0~ (k\sim 0)$ & $ \rm  0.22~(k\sim 0)$ & $\rm 0.99~ (k\sim 0)$\\
\hline 
$\rm $ & $\rm $ & $\rm  $ & $\rm  $ & $\rm  $  &  $ \rm  $\\
$\rm WDOM$ & $\rm 0.0~(k\sim 0.2)$  & $\rm 0.1~(k\sim 0.2)$ & $\rm -0.3~ (k\sim 0.2)$ & $ \rm  0.1~(k\sim 0.2)$ & $\rm 0.79~ (k\sim 0.2)$\\
\hline 
$\rm $ & $\rm $  & $\rm  $ & $\rm  $ & $\rm  $  &  $\rm  $ \\
$\rm NADO1$ & $\rm 0.0~(k\sim 0.4)$  & $\rm 0.1~(k\sim 0.4)$ & $\rm -0.3~ (k\sim 0.4)$ & $ \rm  -0.1 ~(k\sim 0.4)$ & $\rm 0.585~ (k\sim 0.4)$\\
\hline 
$\rm $ & $\rm $  & $\rm  $ & $\rm  $ & $\rm  $   &  $\rm  $ \\
$\rm NADO2$ & $\rm 0.0~(k\sim 0.3)$  & $\rm 0.1~(k\sim 0.3)$ & $\rm 0.0~ (k\sim 0.3)$ & $ \rm  0.2~(k\sim 0.3)$ & $\rm 0.6~ (k\sim 0.3)$\\
\hline

 \hline
 \end{tabular}
\caption{
The above table presents the spin estimates of the BH sources considered in \ref{Table1} from $\chi^2$ minimization. The spin measurements obtained from earlier estimates are also reported.
}
\label{Table4}
\end{center}

\end{table}
To determine the most favored value of $k$ we adopt the following procedure:
\begin{enumerate}
\item We choose a given QPO model as mentioned in \ref{Table2}. 
\item We next choose a source from \ref{Table1}. 
\item Then we select a value of $k$ which automatically fixes the allowed values of spin such that the central singularity is covered by a horizon. We vary the spin in this allowed range.
\item For the chosen $k$ and spin we vary the black hole mass between $(M_{BH}-\Delta M_{BH}) \leq M_{BH} \leq (M_{BH} + \Delta M_{BH})$) where $\Delta M_{BH}$ is the error in the mass measurement (given in \ref{Table1}).
\item Further, for a given combination of $k$, $a$ and $M_{BH}$ we vary $r_{ms}(k,a)\leq r_{\rm em} \leq r_{ms}(k,a) + 20 r_{g}$.
\item The value of $M_{BH}$, $a$ and $r_{\rm em}$ that minimizes the chi-square for the given $k$ are denoted by $M_{\rm m}$, $a_{\rm m}$ and $r_{\rm m}$ for the chosen source. 
\item We repeat steps 3-6 for the other sources keepinf $k$ fixed.
\item Then we repeat steps 3-7 for the remaining values of $k$ keeping the QPO model fixed. This enables us to calculate the magnitude of the joint-$\chi^2$ for the chosen QPO model which we plot as a function of $k$. The magnitude of $k$ where the joint-$\chi^2$ minimizes is denoted by $k_{min}$ and the value of the minimum $\chi^2$ is denoted by $\chi^2_{min}$. The $M_m$ and $a_m$ for each source corresponding to $k_{min}$ are denoted by $M_{min}$ and $a_{min}$ and are reported in \ref{Table3} and \ref{Table4} respectively.
\item Models like RPM can also explain the low-frequency QPO observed in GRO J1655-40, when the form of $\chi^2$ is given by,
\begin{align}
\label{S5-2}
\chi ^2 (k)&=\sum_{j} \frac{\lbrace \nu_{\textrm{u1}{,j}}-\nu_1(k,a_{\rm min},M_{\rm min},r_{\rm em, min}) \rbrace ^2}{\sigma_{\nu_{\rm u1}, j}^2}  
+\sum_{j}  \frac{\lbrace \nu_{\textrm{u2}{,j}}-\nu_2(k,a_{\rm min},M_{\rm min},r_{\rm em,min}) \rbrace ^2}{\sigma_{\nu_{\rm u2}, j}^2} 
\nonumber 
\\
&\hskip 4 cm +\frac{\lbrace \nu_{L,{\rm GRO}}-\nu_3(k,a_{\rm min},M_{\rm min},r_{\rm em,min}) \rbrace ^2}{\sigma_{\nu_{L},\rm GRO}^2}~.
\end{align}
\item We next choose another model from \ref{Table2} and repeat steps 2-9. This gives us the variation of $\chi^2$  with $k$ for each of the QPO models. 
\item When the number of interesting parameters is one, the confidence intervals (i.e., $\Delta \chi^{2}$ from $\chi^{2}_{\rm min}$) associated with 68\%, 90\% and 99\% confidence levels are equal to 1, 2.71 and 6.63 \cite{1976ApJ...210..642A}. The variation of $\chi^2$ with $k$ for each of the QPO models is illustrated in \ref{Fig_07} and \ref{Fig_08}. It is important to note that in the present analysis we have assumed all the black holes to possess similar NED charge parameter $k$. Therefore, when we consider $k$ as the interesting parameter we mean this to be the average charge associated with the black holes.
This assumption is relevant as long as $k$ varies in a very small range which in the present case is $0\lesssim k \lesssim 0.7$ such that taking an average is meaningful. Moreover since $k\geq 0$, if $k=0$ is favored observationally, it implies that the black holes in the sample possess no NED charge.
 Since we are observationally constraining the squared charge to mass ratio $\kappa/r_g$ we do not face difficulties in considering black holes with largely different masses, i.e. if mass increases the charge increases proportionally such that the magnitude of $k$ remains similar and in the range in the range $0\lesssim k \lesssim 0.7$ for the black holes in the sample. 
This allows us to consider a data sample comprising of four stellar mass and one supermassive black hole in this work.

This method is quite general and is not restricted to only regular black holes. The analysis works perfectly fine for regular black holes as well as for black holes with $r=0$ curvature  singularity as long as the metric is characterized by three parameters, namely, mass, spin and some charge.
We have used this method to constrain the tidal charge parameter of braneworld black holes \cite{Banerjee:2021aln} and the magnetic monopole charge parameter of Bardeen black holes where Bardeen black holes are regular black holes with a de Sitter core \cite{Banerjee:2022ffu}. This analysis can be further extended to more general regular black holes.
In each case we estimate the dimensionless charge parameter (which is the charge scaled with the black hole mass) from the observations.

Therefore, method of analysis would not differ if we consider regular black holes with de Sitter cores.
The assumption of Minkowski core comes in the choice of the mass function. For Bardeen black holes or braneworld black holes the mass functions are different. Each time the mass function is characterized by some charge parameter which has some physical interpretation. An external observer perceives the black hole through the spacetime curvature created by the metric parameters, namely, mass, spin and the charge, on which the epicyclic frequencies depend. If the metric comprises of more than three parameters then depending on the situation at hand the $\chi^2$ can be characterized by more than one interesting parameters \cite{1976ApJ...210..642A} in which case we will be required to generate contour plots of $\chi^2$ to estimate the observationally favored magnitude of the charges which quantify the departure from GR.

\end{enumerate}

Although we do not extract the black hole mass from the present analysis we provide independent estimates of spin since previous estimates of spin are derived using \gr\ and different methods of spin measurements assuming \gr\ 
yield largely disparate results (e.g., GRO J1655-40 \cite{Motta:2013wga}). 
In this regard we mention that earlier estimates of spin for GRS 1915+105, GRO J1655-40 and Sgr A* give rise to results with large discrepancies, e.g 
for GRS 1915+105, the Fe-line method gives $a \sim 0.6- 0.98$ \cite{Blum:2009ez} while from Continuum Fitting Method the spin turns out to be intermediate ($a\sim 0.7$ \cite{2006MNRAS.373.1004M}) as well as maximal ($a\sim 0.98$ \cite{McClintock:2006xd}). With revised mass and inclination the spin of GRS 1915+105 turns out to be $0.4<a<0.98$. From \ref{Table4} we note that our results based on PRM and KRM1 are consistent with previous measurements \cite{McClintock:2006xd,Blum:2009ez,Mills:2021dxs}. 
Spin measurements of the source XTE J1550-564 gives $-0.11<a<0.71$ from the Continuum Fitting method while Fe-line method yields a more stringent constraint $a=0.55^{+0.15}_{-0.22}$ \cite{Steiner:2010bt}. Our findings reveal that apart from PRM and KRM1 all the models 
yield spin in the range predicted by the Continuum Fitting method.
Based on the Fe-line method and the Continuum-Fitting method the spin of GRO J1655-40 turns out to be $0.94<a<0.98$ \cite{Miller:2009cw}
and $0.65<a<0.75$ \cite{Shafee_2005} respectively while from QPO data assuming RPM the spin of the source is $a=0.290\pm 0.003$ \cite{Motta:2013wga}. From our analysis we find that our spin estimates based on RPM is in agreement with earlier spin measurements based on QPO related observations. 
Using Continuum-Fitting method \cite{Steiner:2011kd} the spin of H1743-322 turns out to be $a=0.2\pm 0.3 $ (with 68\% confidence) and $a<0.92$ (with 99.7\% confidence). From \ref{Table4} we note that models like RPM, FRM1, KRM3, WDOM, NADO1 and NADO2 yield results in agreement with previous findings. 
For the source Sgr A* the spin is obtained from its radio spectrum as well as motion of S2 stars. Since the radio spectrum of Sgr A* is difficult to model \cite{Reynolds:2013rva,Moscibrodzka:2009gw,Shcherbakov:2010ki} diverse values of spin are obtained, e.g., $a\sim 0.9$ \cite{Moscibrodzka:2009gw}, $a\sim 0.5$ \cite{Shcherbakov:2010ki}. Investigating the motion of S2 stars near Sgr A* reveal the source has a spin $a\lesssim 0.1$ \cite{Fragione:2020khu}. From the study of X-ray light curve of Sgr A* one finds that the object has a maximal spin with ($ \rm a=0.9959\pm 0.0005$) \cite{2010MmSAI..81..319A}. From our analysis we find that models like RPM, KRM3, WDOM  and NADO1 yield a maximal spin as obtained in \cite{2010MmSAI..81..319A}.

In \ref{Fig_07} and \ref{Fig_08} we plot $\chi^{2}$ as a function of $k$ for the QPO models discussed in \ref{S4}. In particular, \ref{Fig_07} depicts the variation of $\chi^2$ with $k$ for (a) Relativistic Precession Model (RPM), (b) Tidal Disruption Model (TDM), (c) Parametric Resonance Model (PRM), (d) 3:1 Forced Resonance Model (FRM1) and (e) 2:1 Forced Resonance Model (FRM2) while \ref{Fig_08} illustrates the variation of $\chi^2$ with $k$ for (a) Keplerian Resonance Model 1 (KRM1) (b) Keplerian Resonance Model 2 (KRM2) (c) Keplerian Resonance Model 3 (KRM3) (d) Warped Disk Oscillation Model (WDOM) (e) Non-axisymmetric Disk Oscillation Model 1 (NADO1) and (f) Non-axisymmetric Disk Oscillation Model 2 (NADO2). The value of $k$ where minimum $\chi^2$ is attained represents the one most favored by observations. We denote this value of $k$ by $k_{min}$. We also plot the confidence intervals, namely, with one interesting parameter, the $1-\sigma$ confidence interval corresponds to $\chi^2=\chi^2_{min}+1$ (plotted with black dashed line), $2-\sigma$ confidence interval corresponds to $\chi^2=\chi^2_{min}+2.71$ (plotted with blue dashed line) while the $3-\sigma$ confidence interval corresponds to $\chi^2=\chi^2_{min}+6.63$ (plotted with magenta dashed line).

From \ref{Fig_07} and \ref{Fig_08} we note that models like Relativistic Precession Model (RPM), Parametric Resonance Model (PRM) and Keplerian Resonance Model 1 (KRM1) establish stringent constraints on $k$. According to RPM, $\chi^2$ attains its minimum value when $k\simeq 0$ and discards $k\gtrsim 0.03$ outside 3-$\sigma$ confidence interval, for PRM $\chi^2_{min}$ is attained at $k\simeq 0.15$ and eliminates $k=0$ outside 2-$\sigma$ confidence interval but includes \gr\ within 3-$\sigma$. In particular PRM eliminates $k\gtrsim 0.3$ outside 3-$\sigma$ confidence interval. 
For KRM1, $k\simeq 0.15$ corresponds to $\chi^2_{min}$ while $k\gtrsim 0.35$ is eliminated outside 3-$\sigma$ confidence interval. 
This is interesting because PRM and RPM which discard large values of the charge parameter $k$ are the most widely used QPO models. In particular, the commensurability of QPO frequencies can be most naturally explained by resonance models like PRM. While KRM1 also puts strong bounds on $k$, it is difficult to realize the resonant couplings assumed in KRM1 in the accretion scenario.

The remaining QPO models do not establish such strong bounds on $k$.
The minimum $\chi^2$ for the remaining QPO models correspond to (i) $k\simeq 0.2$ for the Tidal Disruption Model (TDM) (discards $k\gtrsim 0.65$ outside 3-$\sigma$ confidence interval), (ii) $k\simeq 0.35$ for the 3:1 Forced Resonance Model (discards $k\gtrsim 0.65$ outside 3-$\sigma$ confidence interval), (iii) $k\simeq 0.25$ for the 2:1 Forced Resonance Model (discards $k\gtrsim 0.65$ outside 3-$\sigma$ confidence interval), (iv) $k\simeq 0$ for the Keplerian Resonance Model 2 (discards $k\gtrsim 0.65$ outside 3-$\sigma$ confidence interval), (v) $k\simeq 0$ for the Keplerian Resonance 3 (discards $k\gtrsim 0.6$ outside 3-$\sigma$ confidence interval), (vi) $k\simeq 0.25$ for the Warped Disk Oscillation Model (discards $k\gtrsim 0.65$ outside 3-$\sigma$ confidence interval), (vii) $k\simeq 0.45$ for the Non-axisymmetric Disk Oscillation Model 1 (discards $k\gtrsim 0.6$ outside 3-$\sigma$ confidence interval) and (viii) $k\simeq 0.25$ for Non-axisymmetric Disk Oscillation Model 2.
The above discussion reveals that all the QPO related observations generically discard large values of the charge parameter $k$.




\section{Concluding Remarks}\label{S6}
In this work we study regular black holes with an asymptotically Minkowski core which arise in gravity coupled to non-linear electrodynamics. Such black holes are associated with a non-linear electrodynamics (NED) charge parameter. Study of regular black holes is important as these can evade the curvature singularity at $r=0$, an otherwise unavoidable feature in \gr. Moreover, the exponential convergence factor arising in the mass function makes the quantum gravity model finite to all orders upto the Planck scale. 
Here we aim to decipher the signatures of the NED charge parameter from the observed quasi-periodic oscillations in black holes. Theoretical models aimed to address these peaks in the black hole power spectrum explain them in terms of local or collective motion of plasma near the marginally stable circular orbit. Thus, QPOs turn out to be a cleaner probe to the background spacetime compared to  
the continuum spectrum or the Fe-line which also depend on the complex physics of the accretion flow.

We review the various kinematic and resonant models of QPO proposed in the literature and compare the model based QPO frequencies with the available observations. In particular, we evaluate error estimator like the $\chi^2$ which enables us to derive the observationally favored charge parameter and the spin corresponding to each of the black holes. Our analysis reveal that most QPO models favor the general relativistic scenario or small values of $k$. Particularly, models like RPM, PRM and KRM1 establish stringent constraints on the NED charge parameter, e.g, PRM and KRM1 favor small but non-trivial values of the charge parameter, viz, $k\simeq 0.15$ but eliminates large values of $k$ e.g. $k\gtrsim 0.35$ outside 3-$\sigma$ confidence interval. The Relativistic Precession model on the other hand favor $k=0$ and discard $k\gtrsim 0.03$ outside 99.7\% confidence level. This is interesting because the most extensively used QPO models like RPM and PRM impose very strong bounds on the NED charge parameter. We further note that all the QPO models studied here discard large values of the charge parameter. It is important to note that the observationally preferred magnitude of $k$ corresponds to the average charge of the black holes and in the present analysis we assume that the black holes considered here have similar charges. A separate analysis considering black holes with different charges can be done in which case the number of interesting parameters will be two i.e., charge and spin and one needs to consider the sources individually to obtain constraints on the observationally favored magnitude of these two parameters. We leave this analysis for a future work.

Although it might seem from the present analysis that large values of the non-linear electrodynamics charge parameter are disfavored by observations related to QPOs, these results have certain limitations. First, our findings are purely model dependent, i.e., despite observing QPOs for decades there is no common concensus on the correct choice of the QPO model. While some believe that HFQPOs in black holes should be explained by a single QPO model which is yet to be determined, others believe that the correct choice of the QPO model may be source dependent. Second, there are very few black holes which exhibit HFQPOs in their power spectrum, and hence our results are limited due to poor statistics. Moreover, the present data has large errors, which further limits our analysis. This can however be overcome with the launch of the ESA (European Space Agency) X-ray mission LOFT (Large Observatory for X-ray Timing) which aims to improve the precision of data by an order of magnitude. However, with the available data sample, precision and models, the present analysis provides a possible framework which may be used to constrain several alternatives to \gr, with QPO related observations.

\section*{Acknowledgement}
Research of I.B. is funded by the Start-Up
Research Grant from SERB, DST, Government of India
(Reg. No. SRG/2021/000418).


\bibliography{regularBh,bardeen,new-ref,KN-ED,QPO,Brane,IB,Gravity_3_partial,Gravity_1_full,Black_Hole_Shadow,EMDA-Jet,QG}

\bibliographystyle{./utphys1}
\end{document}